# VLTI/AMBER differential interferometry of the broad-line region of the quasar 3C273[*]

Romain G. Petrov[†a], Florentin. Millour[a], Stéphane Lagarde[a], Martin Vannier[a], Suvendu Rakshit[a], Alessandro Marconi[b] and Gerd Weigelt[c]

[a]Laboratoire Lagrange, UMR 7293, University of Nice Sophia-Antipolis, CNRS, Observatoire de la Côte d'Azur, BP 4229, 06304 Nice Cedex 4, France; [b]Department of Physics and Astronomy, University of Florence, Largo Enrico Fermi 2, 50125, Firenze, Italy; [c]Max Planck Institute für Radioastronomie, Auf dem Hügel 69, 53121, Bonn, Germany.

**ABSTRACT**

Unveiling the structure of the Broad-Line Region (BLR) of AGNs is critical to understand the quasar phenomenon. Resolving a few BLRs by optical interferometry will bring decisive information to confront, complement and calibrate the reverberation mapping technique, basis of the mass-luminosity relation in quasars. BLRs are much smaller than the angular resolution of the VLT and Keck interferometers and they can be resolved only by differential interferometry very accurate measurements of differential visibility and phase as a function of wavelength. The latter yields the photocenter variation with λ, and constrains the size, position and velocity law of various regions of the BLR. AGNs are below the magnitude limit for spectrally resolved interferometry set by currently available fringe trackers. A new "blind" observation method and a data processing based on the accumulation of 2D Fourier power and cross spectra permitted us to obtain the first spectrally resolved interferometric observation of a BLR, on the K=10 quasar 3C273. A careful bias analysis is still in progress, but we report strong evidence that, as the baseline increases, the differential visibility decreases in the $Pa_\alpha$ line. Combined with a differential phase smaller than 3°, this yields an angular equivalent radius of the BLR larger than 0.4 milliarcseconds, or 1000 light days at the distance of 3C273, much larger than the reverberation mapping radius of 300 light days. Explaining the coexistence of these two different sizes, and possibly structures and mechanisms, implies very new insights into the BLR of 3C273.

## 1. INTRODUCTION

### 1.1 QSOs and their BLR

Quasars and Active Galactic Nuclei (AGN) are extremely bright sources almost certainly powered by accretion onto a central super massive black hole (SMBH). They dominate the night sky in many wavelength domains and radiate 1/5 of the power in the Universe. They are important to the evolution of their galaxy and, as they can be observed at very high redshifts, they are tags to the global evolution of the Cosmos. They are the seed of extreme physical processes and one of the main targets of relativistic astrophysics.

The main components of an AGN are the central SMBH surrounded by a very compact accretion disk emitting mainly in the continuum. The nucleus is surrounded by a dust torus that obscures the central region in type 2 AGNs[1]. Some of the material inflowing from the galaxy is eventually ejected, and might contribute to the high-velocity jets.

---

[†] Contact: romain.petrov@unice.fr; phone +33 04 92 00 39 61; lagrange.oca.eu.

Type 1 Active Galactic Nuclei are not obscured by the dust torus and show broad emission lines (and sometimes absorption lines), with widths of several thousands of km/s. This broad line region (BLR), sometimes called the "atmosphere" of the quasar[2], is the crossroad of the inflow and outflow of material. Understanding the broad line regions "is of critical importance to understand the quasar phenomenon: (1) to understand how the accretion/outflow processes work in AGNs and (2) to understand the geometry and kinematics of the BLR to correct the AGN black hole mass measurement"[3]. So far, the morphological information about BLRs is given by reverberation mapping (RM), which studies the delay needed for an intensity variation generated near the accretion disk in the continuum to be echoed by the different velocity bins of an emission line. This yields an echo diagram $\psi(V,t)$ that in theory can constrain very strongly the morphology of the BLR. In practice, we can generally measure only the line profile ($L(V) = \int \psi(V,t)dt$), which yields a radial velocity amplitude $\Delta V$ and the mean delay $\tau$ between the continuum flux variation and the global response of a central part of the emission line. The delay $\tau$ is related to the equivalent width of the delay transfer function $\int \psi(V,t)dt$. These two measures yield an estimate of the black hole mass, taken to be:

$$M_{BH} = fr\Delta V^2/G \qquad (1)$$

where $r = \tau c$ is the reverberation mapping radius of the BLR. The projection factor $f$, the radius $r$ and the line width $\Delta V$ depend on the exact morphology and kinematics of the BLR, and the resulting uncertainty on the BH mass can exceed a factor three[3], with important consequences on the broadly used general mass-luminosity relationship for QSOs.

**1.2 Interferometric observations of AGN dust tori**

After pioneering works by Jaffe[4] and Swain[5] in 2004, a handful of Sy1 AGNs have been successfully observed in the last two years by optical interferometry in low spectral resolution in the K band, with the Keck Interferometer[6] and with the VLTI[7]. The broadband absolute visibility in the K band is fairly high but reveals a partial resolution of the AGN image in the near-infrared continuum. There are several contributions to the continuum spectra: the dust torus, probably dominated by its hotter part that is at the inner rim; the central very compact accretion disk, the gas in the BLR itself and the synchrotron emission. From theory, spectral energy distribution, light curves and polarizations considerations, Kishimoto[6,8] evaluates that the K band continuum image is dominated by the ring-like inner rim of the dust torus and derives its equivalent radius. Higher spectral resolution observations in an emission line can select the contribution of the gas producing the BLR and constrain its position and angular size, as discussed below.

**1.3 Differential interferometry of BLRs**

The largest reverberation mapping radii reach a few hundreds of light days. At the distance of the nearest Quasars, this corresponds to angular sizes of the order of typically 100 μas, well below the resolution limit of any current optical interferometer.

It is possible to obtain angular information on non-resolved sources by measuring the displacement of their photocenter with wavelength. This has been proposed first in the context of speckle interferometry[9], extended to long baseline interferometry[10] and has been one of the design parameters of the VLTI focal instrument AMBER[11], which has given several results with an accuracy better than 30 μas[12]. For sources much smaller than the resolution limit λ/B, a photocenter displacement $\varepsilon(\lambda)$ in the direction of the baseline $B$ will produce a differential phase $\phi(\lambda)$ given by[10]:

$$\phi(\lambda) = 2\pi \frac{B}{\lambda} \varepsilon(\lambda) \qquad (2)$$

If $\alpha$ is the equivalent angular size of the source, the closure phase $\Psi(\lambda)$ decreases as $\alpha^3$, the visibility drop $1 - V(\lambda)$ decreases as $\alpha^2$, and the amplitude of the differential phase $\phi(\lambda)$ decreases as $\alpha$, which makes it the most useful observable to seek angular information on non resolved targets observed at high SNR[13]. In this paper we will discuss and illustrate the method on the quasar 3C273. Figure 1 displays the expected velocity field of a Keplerian disk BLR and the corresponding photocenter and differential phase variations[14], observed through the Paschen α line in spectral bins of 400 km/s. For the quasar 3C273 the expected photocenter displacement is of the order of 20 μas, i.e. about 0.5% of the interferometer resolution. Measuring it requires an accumulated SNR=200 on the coherent flux. With AMBER[11] on the VLTI with the UTs, this can be achieved in typically 10 to 20 minutes if we are fringe tracking and 2 or 3 hours if we have to observe in the "blind mode" described below. Measuring a photocenter displacement will constrain the geometry of the BLR. For example, a Keplerian disk and a virialized 3D distribution of clouds with the same reverberation mapping radius $r$ and line width $\Delta V$ have very different photocenter displacements. The photocenter $\vec{\varepsilon}(\lambda)$ for a Keplerian disk is illustrated in figure 1. The 3D distribution will give zero photocenter displacement. Combining photocenter displacement measures with the reverberation mapping information should allow us to identify a model for the BLR structure and velocity field. Then, photocenter measures yield the velocity as a function of radius function and hence the mass of the SMBH. It will allow a more precise estimate of the RM projection factor $f$, a better separation between global and local velocity field effects and then a better measure of $\Delta V$. This will constrain the mass of the observed quasar and of all quasars with similar RM properties. In addition, the combination of the angular photocenter measure with the linear RM radius yields a direct estimate of quasar distance.

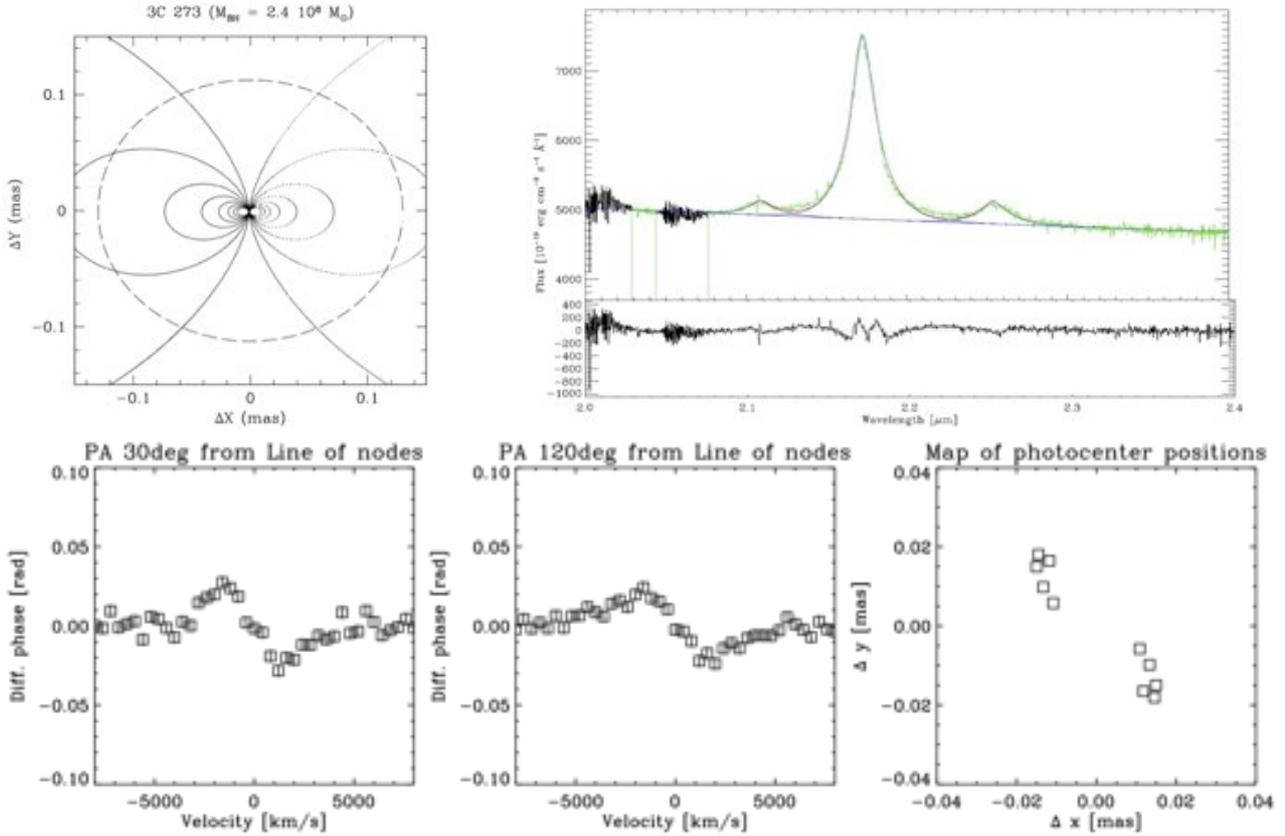

Figure 1: Evaluation of the photocenter displacement and differential phase for a Keplerian disc BLR in 3C273. Top left: equal radial velocity curves in a Keplerian disk with 30° inclination. The curves delimitate 400 km/s bins. The photocenter of the BLR in each bin is reduced by the contribution of the continuum, indicated by the measured $Pa_\alpha$ line profile (top-right). The resulting photocenter, in the bottom right figure, has amplitude of $\pm 20 \mu as$. With a 100m baseline this gives phase amplitude of up to 0.05 radians (2.5°) depending of the angle between the baseline and the photocenter displacement direction, as illustrated in the bottom-left and bottom-center figures. The error boxes have a size of 2°.

## 1.4 3C273

3C273 is the brightest nearby quasar and one of the "brightest known sources in the Universe". It is red-shifted enough for the $Pa_\alpha$ line to be in the K band and this makes it particularly suited for these observations because the emission is at least two times stronger in $Pa_\alpha$ than in $Br_\gamma$. The main characteristics of 3C273 are:

- Magnitude: K=9.7
- Red shift: z=0.16
- Paschen α line center: 2.17 µm
- Paschen α line width: $\Delta V = 3400 \ km/s$
- Reverberation mapping radius, in $H_\gamma$: $307^{+69}_{-91}$ corresponding to 95 µas[15]
- Reverberation mapping radius, in $H_\alpha$: $514^{+65}_{-64}$ corresponding to about 160 µas[15]

- Absolute visibility in the K band (Keck Interferometer, 84 m projected baseline)[8]: 0.978 ± 0.017
- Radius of the inner rim of torus: 0.81 ± 0.34 pc corresponding to 0.296 ± 0.124 mas[8]

## 2. BLIND MODE OBSERVATIONS AND 2DFT DATA PROCESSING

To study the BLR we need to resolve spectrally the emission line. A spectral resolution of 200 gives a few points in the line and global constraints on the BLR position and size. A resolution higher than 500 gives access to more than 10 velocity bins and to a fit of the global velocity law. In the near infrared, optimized observations at such resolutions imply the use of a fringe tracker to stabilize the fringes and allow exposure times longer than the piston coherence time, necessary to get out of the detector noise regime. When we first proposed differential interferometry observations of BLRs with the AMBER/VLTI instrument[11], we assumed fringe tracking and our "conservative" estimation for photocenter accuracy was of about 2 µas in 1 hour of observation for K=10. This would allow to apply the technique to two dozens of targets.

The ESO call for proposals[16] for the AMBER/VLTI instrument offers medium resolution observations only at magnitude permitting fringe tracking, which is 7.5 with the UTs in the CfP and can approach 8.5 in very good conditions. The severe limitation of FT limiting magnitude seems a general situation. The reason is that current fringe trackers need a sufficient SNR in extremely short exposures, to be able to freeze the piston at a fraction of wavelength. The consequence is that it will be very difficult to have fringe trackers allowing the observations of AGNs in medium spectral resolution.

Actually, it is not necessary to detect fringes in each individual frame. With the medium spectral resolution of AMBER, the coherence length in the K band is of about 3 mm. The atmospheric piston jitter has a RMS amplitude of typically a few tens of microns and the delay line model errors and drift are below 100 µm/mn. So, after centering the fringes on a bright calibrator, we have at least half an hour to observe a faint target with the guarantee that the fringes are present in the data, even if each individual frame looks just as detector noise. We still have to make exposures short enough to have a good contrast of the fringes. We have to integrate values which are not sensitive to the piston value but still contain information about the source visibility, differential phase and closure phase.

### 2.1 2DFT fringe detection

All the interferometric quantities that can be averaged in spite of an unknown and variable phase in individual interferograms can be used to reduce data recorded in blind mode. This includes the modulus of Fourier interferograms, yielding source visibility, the Fourier bi-spectrum, yielding closure phase, and cross spectra between different spectral channels yielding the chromatic differential phase and visibility. We have chosen to use the 2D Fourier transform of x-λ interferograms, similar to this used in REGAIN/GI2T and later VEGA/CHARA[17]. It has the advantage to allow a straightforward and unambiguous detection of the average group delay after some integration time and is therefore easy to use as a low frequency coherencing sensor correcting slow drifts of the OPD. It has also some similarity with the tool that we use to analyze the dark current and sky images on our detector and correct our data from detector fringes[18].

Let's $i_m(x, \lambda)$ be the AMBER x-λ image, with its 3 dispersed fringe patterns, illustrated in figure 2a. As a first step, we resample the data to obtain the interferogram $i(x, \sigma)$ where the fringes in all spectral channels have the same interfringe (corresponding to an average wavelength $\bar{\lambda} = 1/\bar{\sigma}$) and the spectral channels are equally spaced in wavenumber σ=1/λ. This is done by a bilinear interpolation and should have a limited impact on the quality of the data since our frames are substantially oversampled both in spatial and spectral directions. The Fourier transform of $i(x, \sigma)$ in each spectral channel yields the 1d Fourier interferogram $I(u, \sigma)$ which can be written as :

$$I(u,\sigma) = \mathcal{F}_x[i(x,\sigma)] = n(\sigma)F(u,\sigma)\sum_i n_i + \sum_{i,j>i} \sqrt{n_i n_j}\, \Omega(u,\sigma) e^{2i\pi\sigma p_a^{ij}} \qquad (3)$$

where $n_i$ is the total contribution of telescope i to the number of photons in the interferogram ; $n(\sigma)$ is the source spectrum as seen by the instrument ; $F(u, \sigma)$ is the Fourier transform of the resampled interferometric window, $p_a^{ij}$ is the achromatic part of the piston difference on the baseline i-j and $\Omega(u, \sigma)$ is the Fourier transform of the source seen by the instrument, defined by :

$$\Omega(u,\sigma) = n(\sigma) V_I(u,\sigma) V_*(u,\sigma) \exp\left[ i\phi_*(u,\sigma) + i\phi_I(u,\sigma) + 2i\sigma p_c^{ij}(\sigma) \right] * F(u,\sigma) \qquad (4)$$

where $V_I(u,\sigma)$ and $V_*(u,\sigma)$ are the instrument and source visibility at the spatial frequency u and wavenumber σ; $\phi_I(u,\sigma)$ and $\phi_*(u,\sigma)$ are the corresponding instrument and source phase and $p_c^{ij}(\sigma)$ is the chromatic part of the piston difference :

$$p^{ij}(\sigma) = p_i(\sigma) - p_j(\sigma) = p_a^{ij} + p_c^{ij}(\sigma) \qquad (5)$$

where the achromatic OPD difference $p_a^{ij}$ corresponds to what is usually called the « piston » difference between beams i and j and $p_c^{ij}(\sigma)$ contains all the wavelength dependent terms of the OPD, which are dominated by the dispersion in the VLTI tunnels and AMBER fibers. $p_a^{ij}$ very rapidly varies with time while $p_c^{ij}(\sigma)$ is dominated by terms evolving more slowly as the source zenith distance changes. To simplify the equations in the following, we shall assume that the window function $f(x,\sigma)$ is flat and with constant size and hence $F(u,\sigma) = \delta(u)$.

A Fourier transform of the interferogram in equation (3) in the wavenumber direction yields the 2D Fourier transform :

$$\hat{I}(u,v) = \mathcal{F}_\sigma[I(u,\sigma)] = \hat{n}(v) * \hat{F}(u,v) \sum_i n_i + \sum_{i,j>i} \sqrt{n_i n_j}\, \hat{\Omega}(u,v) * \delta(v - p_a^{ij}) \qquad (6)$$

and its average power spectrum :

$$D(u,v) = <|\hat{I}(u,v)|^2> = |\hat{n}(v) * \hat{F}(u,v)|^2 \sum_i n_i + \sum_{i,j>i} \sqrt{n_i n_j} |\hat{\Omega}(u,v)|^2 * <\delta(v - p_a^{ij})> \qquad (7)$$

$D(u,v)$ displays a low frequency peak and one fringe peak for each baseline at the position $u = B_{ij}\bar{\sigma}$ and $v = p_a^{ij}$ as illustrated in figure 2d.

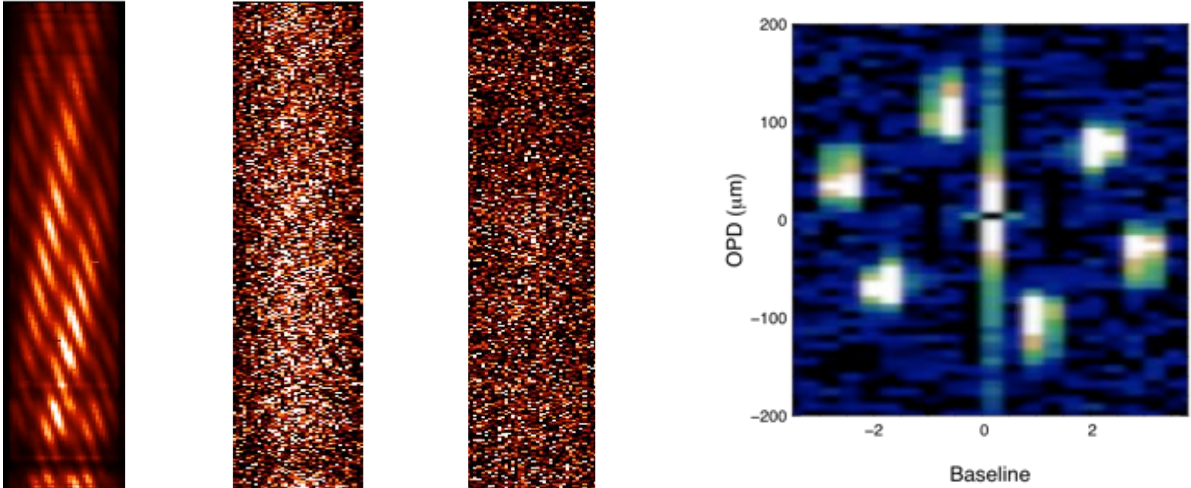

Figure 2: Principle of fringe detection in blind mode observations. The 3 left figures are x-λ interferograms with fringes dispersed in the vertical direction. The magnitudes are K=4 (left), K=8.5 (center) and K=9.7 (right). The rightmost figure represents a 10' average 2D FT of the interferograms for K=9.7.

The fringe peak will grow with the number of frames, as long as it moves by less than its size, while the noise level will grow only like the square root as the number of frames. The typical size of the fringe peak is given by the spectral coverage of the initial interferogram. For AMBER in medium resolution this is between $\lambda_1 = 2\mu m$ and $\lambda_1 = 2.3\mu m$, yielding a fringe peak width of $\lambda_1\lambda_2/(\lambda_2 - \lambda_1) = \lambda^2/\Delta\lambda = 15\mu m$. Under standard conditions, it takes at least a few seconds for the piston to drift by that value, and this sets a limit of the technique: we can observe sources producing a fringe peak of sufficient SNR in a few seconds to allow a piston measurement and correction. If the classical limit of the instrument is set by the necessity to detect SNR=3 fringes in say 100 ms and we consider that blind observing can manage the same SNR=3 criteria over say 10s, then we can afford a fringe peak SNR per frame of 0.3. In detector noise regime, this corresponds to a source 10 times fainter and hence a gain of 2.5 magnitudes. The gain is even more important with regard to a fringe tracker that must reach an SNR of the order of 3 in much shorter frame times.

In Figure 2a, we display a standard AMBER x-λ individual interferogram, for a K=4 bright source. The 3 fringe systems are clearly seen. Figure 2b shows a K=8.5 interferogram. Fringes are quite hard to see, but frame-by-frame data processing detects fringes and measures a piston in at least in some frames. This is the limit of the standard AMBER data processing. In figure 2c, with K=9.7, any frame-by-frame processing fails. The 2DFT processing yields the average power spectrum displayed in figure 2d with 3 clear fringe peaks. Piston offsets have been introduced to clearly separate the fringe peaks. The peak blur corresponds to the piston drift in 10 mn. We see here that it is smaller than 50 μm.

The position of the fringe peak in the 2DFT modulus can be used to evaluate and correct the piston value. Figure 3 displays the cuts of figure 2d in the piston direction at the frequency of each baseline, as they evolve in time. The position of the fringe peak yields the absolute piston (group delay) evolution with time.

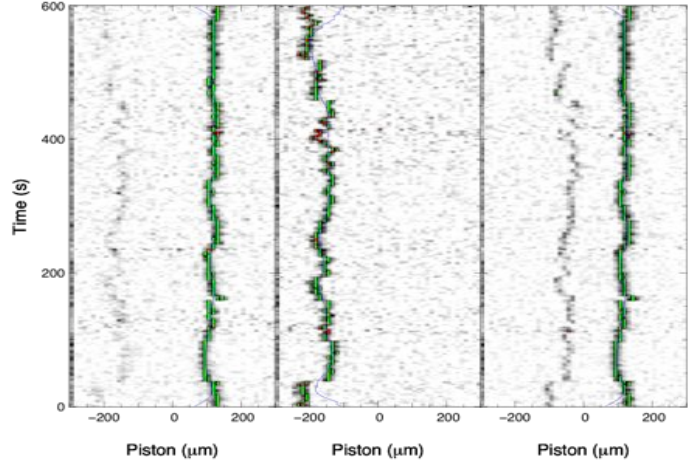

Figure 3: Measurement of piston (OPD group delay) by 2DFT processing on 3C273. Each plot shows a cut of figure 2d at the fringe peak frequency, in the piston direction (here x), repeated over time (here y). The fringe peak is green when it is more than 3 times higher than the background noise. The power spectra are averaged over 1s. We see that we are still not at the sensitivity limit. Note than on baselines 1-2 (left) and 1-3 (right) the OPD remains stable, within a 15 μm fringe peak, for very long periods of time. This plot also allows estimations of the rms error on the fringe peak position, of the order of 15 μm for baseline 1-2 to about 50 μm for baseline 2-3.

## 2.2 Differential observables in 2DFT data processing

When we observe in blind mode and plan to reduce the data with 2DFT algorithm, piston offsets ensure that the fringe peaks are well separated in the piston direction and cannot overlap, solving the problem of baseline cross-talk common to AMBER and other all-in-one multi-axial beam combiners. These offsets are apparent in figure 2d. They are of the order of 100 μm, which remains very small with regard to the coherence length of 3 mm. As the fringe peaks are well separated, we will say that we process them individually and in equations (3) and (4) we will use only the part specific to each baseline. The new interferogram for the baseline $ij$ is:

$$I^{ij}(\sigma) = \sqrt{n_i n_j}\, \Omega^{ij}(\sigma)\, e^{2i\pi\sigma p_a^{ij}} \quad (8)$$

with
$$\Omega^{ij}(\sigma) = \Omega(u = B_{ij}\, \bar{\sigma}, \sigma) = n(\sigma) V_l^{ij}(\sigma) V_*^{ij}(\sigma) \exp\left[ i\phi_l^{ij}(\sigma) + i\phi_*^{ij}(\sigma) + 2i\pi\sigma p_c^{ij}(\sigma) \right] \quad (9)$$

The estimation of the differential measurables at the frequency $u_{ij}$, is based on the computation, for each σ, of the differential cross spectrum (DCS) $W^{ij}(\sigma)$ between a 2D interferogram in which all channels but the channel σ have been forced to zero and a 2D interferogram in which only the channel σ has been forced to zero[‡]:

$$W_\sigma^{ij}(\nu) = \mathcal{F}[I^{ij}(\sigma').\delta(\sigma' - \sigma)].\mathcal{F}\left[I^{ij}(\sigma').\left(1 - \delta(\sigma' - \sigma)\right)\right]^*$$
$$= \sqrt{n_i n_j}\Omega^{ij}(\sigma)\left[\widehat{\Omega^{ij}}\left(\nu - p_a^{ij}\right) - \Omega^{ij}(\sigma)\right]e^{-2i\pi\sigma(\nu - p_a^{ij})} \quad (10)$$

If we know exactly the achromatic piston $p_a^{ij}$ from 2DFT power spectrum, we have

$$W_\sigma^{ij}(\nu = p_a^{ij}) = \sqrt{n_i n_j}\,\Omega^{ij}(\sigma)\big[\widehat{\Omega^{ij}}(0) - \Omega^{ij}(\sigma)\big] = \sqrt{n_i n_j}\,\Omega^{ij}(\sigma)\big[\int \Omega^{ij}(\sigma')\,d\sigma' - \Omega^{ij}(\sigma)\big] \quad (11)$$

---
[‡] To make sure that the DCS does not contain power spectrum terms affected by a quadratic bias.

The term $R(\sigma) = \int \Omega^{ij}(\sigma')\,d\sigma' - \Omega^{ij}(\sigma)$ is very close to be constant. For the simplicity of the equations above, we will write $R(\sigma) \simeq R$, even if the actual $R(\sigma)$ is used in the practical data processing. If we assume that $n_i = n_j = n$ and $V_I^{ij}(\sigma) = V_I$, we see that $|W_\sigma^{ij}(p_a^{ij})| \simeq n(\sigma) V_I^{ij}(\sigma) n^2 V_I^2$ : the DCS is proportional to the square of the flux and the square of the instrumental visibility but its *variations* with σ are proportional to the source differential visibility.

From the measures in the photometric channel of the spectrum $n(\sigma)$ and the fluxes $n_i$ and $n_j$ and from the DCS on the science and the reference sources we build :

$$E^{ij}(\sigma) = \frac{w_{\sigma*}^{ij}(p_{a*}^{ij})/n_*(\sigma)\sqrt{n_i n_j}}{w_{\sigma cal}^{ij}(p_{acal}^{ij})/n_{*cal}(\sigma)\sqrt{n_{ical} n_{jcal}}} = \frac{\Omega_*^{ij}(\sigma) R_*}{\Omega_{cal}^{ij}(\sigma) R_{cal}} \quad (12)$$

Assuming that the instrumental visibility and phase are the same for the science and the reference, we get :

$$E^{ij}(\sigma) = V_*^{ij}(\sigma) \exp\{i\phi_*^{ij}(\sigma) + 2i\pi\sigma\,[p_{c*}^{ij}(\sigma) - p_{ccal}^{ij}(\sigma)]\}\,\frac{R_*}{R_{cal}} \quad (13)$$

To avoid errors in the calibration of the ratio $R_*/R_{cal}$, for example from changes in the instrumental visibility, we divide $E^{ij}(\sigma)$ by its average over σ and we get finally the estimators of differential visibility and phase :

$$\phi_{d*}^{ij}(\sigma) = \arg\left(\frac{E^{ij}(\sigma)}{<E^{ij}(\sigma)>_\sigma}\right) = \phi_*^{ij}(\sigma) + 2\pi\sigma\Delta p_c(\sigma) \quad (14)$$

$$V_{d*}^{ij}(\sigma) = \frac{\Re\big[E^{ij}(\sigma)e^{-i\phi_{d*}^{ij}(\sigma)}\big]}{<\Re\big[E^{ij}(\sigma)e^{-i\phi_{d*}^{ij}(\sigma)}\big]>_\sigma} = \frac{V_*^{ij}(\sigma)}{<V_*^{ij}(\sigma)>_\sigma} \quad (15)$$

The term $2\pi\sigma\Delta p_c(\sigma)$ is the change in chromatic dispersion between science and calibrator. These two terms can be minimized by a correction of the computed chromatic OPD. Since we are looking for sharp differential phase variations through the emission line, we simply perform a polynomial fit of the chromatic OPD and the phase offset for all spectral channels outside the spectral line.

### 3. OBSERVATIONS AND DATA PROCESSING

We observed the quasar 3C273 in May 2011 with AMBER in medium resolution and the UTs 1,2 and 4. We used frame times of 300 ms, and collected about 200 photons per channel and per frame. This represents about 3 photons per pixel and is well below the detector read out noise of 11e[-]. The seeing conditions were very good, from 0.5 to 0.8 arcseconds, stable between 0.6 and 0.7 most of the time. During the half observing night presented here, we collected 47 exposures on 3C273, each with 200 frames of 300 ms. That is 1h20' of open shutter time on 3C273 but only 47' of actual integration. In addition, we recorded a collection of calibrators of different magnitudes. In spite of the fairly long DIT, the VLTI/AMBER visibility was between 0.2 and 0.4, depending on the baseline and the conditions.

### 3.1 Calibrator differential visibilities

Figure 4 illustrates the data processing on a set of calibrators. Figure 4a shows the measured spectrum $n(\sigma)$, 4b shows the measured cross spectrum $W_\sigma^{ij}(p_a^{ij})$ and figure 4c shows the calibrator differential visibility. We used 4 calibrators with magnitudes K=6.6, K=9, K=8.2 and K=9. We see that the behavior of the differential visibility is very stable and can be calibrated with an accuracy better than 1% on the K=9 target. We observe that the differential visibility is bent, mainly because the effect of chromatic OPD has not been corrected in this « first » data processing. The spectral 508 channels obtained in the AMBER medium resolution observations have been binned by groups of 16 for SNR reasons (on the science target) and we have now 0.009 μm per channel, corresponding to a resolution R=240.

## 3.2 3C273 differential visibilities

Figure 5 shows results on 3C273. In figure 5a we see the spectrum of 3C273. We note the same telluric and instrumental lines as for the calibrators, at 2.01 and 2.06 μm and the Paα emission line red-shifted at 2.17 μm. Figure 5b and 5c show the differential cross spectrum and the differential visibility for a 50 m baseline (5b) and 125 m (5c). The emission line appears very clearly in the DCS in figure 5b while it is quite erased in the DCS in figure 5c. This indicates a differential visibility decrease in the line when the baseline increases, which can be seen in the differential visibility plots. However, figure 5 also shows a flux dependent bias of the DCS, which strongly affects the differential visibility in the telluric lines and casts suspicion on the variation in the emission line. In addition, the general shape of the differential visibility in the continuum is far from 1 and changes with the baseline. A bias analysis is needed before confirming the differential visibility measure in the line.

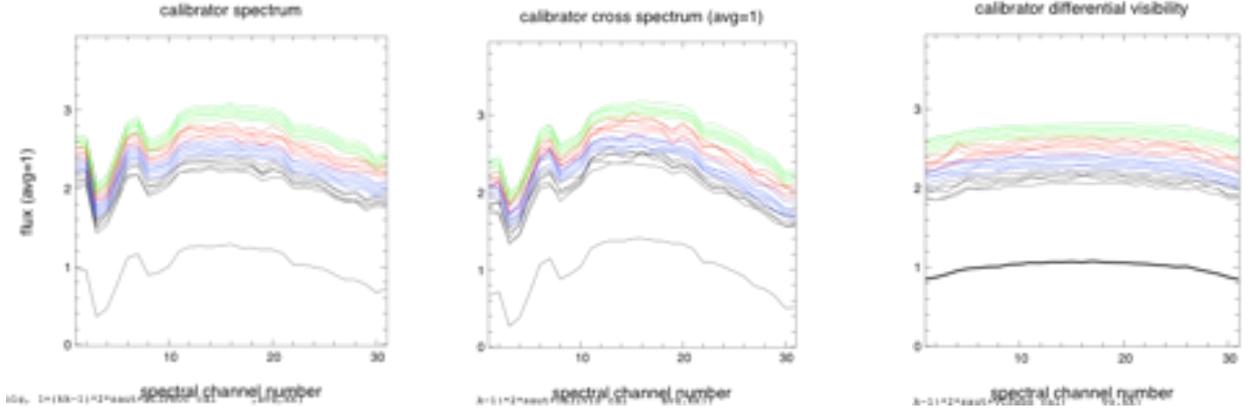

Figure 4: differential visibility measured on a calibrator. Figure 4a (left): calibrator spectrum $n(\sigma)$. Figure 4b (center): calibrator DCS $|W_\sigma^{ij}(p_a^{ij})| \simeq n(\sigma)V_I^{ij}(\sigma)n^2V_I^2$. Figure 4c (right): calibrator differential visibility $V_{d*}^{ij}(\sigma)$. All function are divided by their avergage value over σ. They are shifted in the plot for visibility. The wavelength range goes from 1.99 to 2.31 μm. The black curve around 1 represents the time average. The color curves represent 4 different calibrators. From top to bottom the calibrator magnitudes where K=6.6 (green), K=9 (red), K=8.2 (blue) and K=9 (black). All curves are plotted for the baseline UT1-UT4=125m.

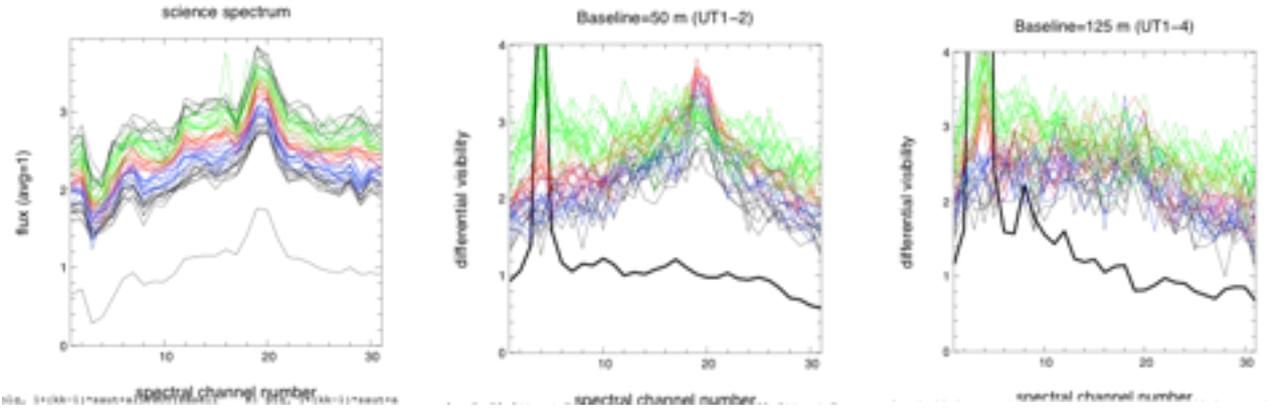

Figure 5: differential cross spectrum and differential visibility on 3C273. Figure 5a (left): 3C273 spectrum. Figure 5b and 5c: differential cross spectrum (top, thin, color curves) and differential visibility (bottom, thick, black curves). Figure 5b (center) is for the 50m UT1-2 baseline. Figure 5c (right) is for the 125m UT1-4 baseline.

## 3.3 Differential phases

Figure 6 shows the differential phases obtained on the calibrators and on 3C273. On the calibrators the differential phase displays the expected differential chromatic OPD. The 3C273 differential phases are always flat. The 3C273 observations were affected by the same chromatic OPDs, since 3C273 and all calibrators were less than 2° apart, and they were interlaced in time. This reveals a bias on the differential phase at the faintest magnitudes.

### 3.4 Bias analysis

Both the differential visibility and phase show biases on 3C273 that do not appear or appear only marginally on the faintest calibrators. The differential cross-spectral measures are not affected by the "quadratic noise bias" that appears in power spectra containing terms of squared zero mean noise.

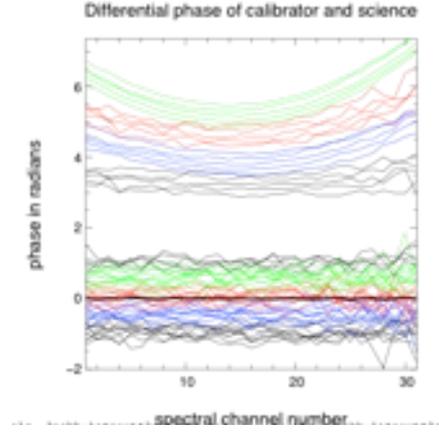

Figure 6: Differential phases in radians. All curves are at 0 average and shifted for visualization. The top curves, between « 3 and 7 rad » are for the same exposures on calibrators than in figure 4. The bottom curves represent the differential phases on the 3C273 exposures described for figure 5. The thick black line represents the average differential phase on 3C273. The color codes in 3C273 and in the calibrators are matched in time: we obtained first the black curves on 3C273, then the black curves on a calibrator, then the blue curves on the science followed by the blue curves on a calibrator and so on.

On 3C273, the DCS is overestimated in the lowest parts of the telluric lines. A careful look at the calibrator plots show that some of the exposures on the faintest calibrator are slightly biased in the bottom of the 2.01 line. The K=9 calibrator is fainter in this line than 3C273 in the continuum around the emission line and in first approximation we should say that the 3C273 measures are not biased around the emission line. We have used an estimate of this bias as a function of the correlated flux made in all available channels outside the emission line. It reduces, but does not cancel, the bias in the telluric line and does not change much the visibility variation in the emission line. This does not really correct the general shape of the differential visibility and differential phase of 3C273 in the continuum, which is mainly due to the "piston error" bias analyzed in the next section.

### 3.5 « Piston error » bias

Our differential visibility estimator in equation (11) assumes that we have an accurate estimate of the achromatic piston $p_a^{ij}$. If we make an error $\Delta p$ on the estimation of the piston, the differential cross spectrum becomes

$$W_\sigma^{ij} = n_i n_j \Omega^{ij}(\sigma) < \left[\widehat{\Omega^{ij}}(\Delta p) - \Omega^{ij}(\sigma)\right] e^{-2i\pi\sigma(\Delta p)} > \tag{16}$$

and our measures are affected by a bias term

$$B(\sigma) = < \left[\widehat{\Omega^{ij}}(\Delta p) - \Omega^{ij}(\sigma)\right] e^{-2i\pi\sigma(\Delta p)} > \tag{17}$$

This term cannot be corrected using the calibrator, since the piston error depends on the source magnitude and on the observation conditions. $B(\sigma)$ depends on the source visibility, on the variations of instrument visibility with σ and on the chromatic OPD, since all these effects will change $\widehat{\Omega^{ij}}(\Delta p)$. In particular, the biases on the DCS detected in the telluric lines will behave like strong and random variations of $V_I^{ij}(\sigma)$ and hence of $\widehat{\Omega^{ij}}(\Delta p)$. Figure 7 displays a Monte Carlo simulation of the effect of $B(\sigma)$ on the differential visibility and phase. We have simulated a target with a flat visibility, but for a sharp local variation of 10% at the position of the emission line, and a zero differential phase but for a sharp local variation of 0.05 radians. The object is affected by window and gain table correction errors, which bend the overall differential visibility, and by variable local biases that mimic the behavior of the DCS the telluric lines. The differential phase is affected by a variation of the chromatic OPD similar to the one observed on the calibrators through the 3C273 observations. Figure 7a displays the evolution of the real differential visibility and phase through the observations. To evaluate the bias term, we have generated one random piston error per frame, with a standard deviation of 5 μm in figure 7b (good calibrator case) and of 50 μm in figure 7c (worst 3C273 case). Each plot in figure 7b and 7c represents an exposure of 200 frames. We see that with a small piston error, the bias has almost no effect on the differential measures. This is coherent with the good measurements on the calibrators. With a strong piston error, both the differential visibility and phase are severely biased, and their error bars are increased. The average broadband visibility is substantially modified. However the sharp variations in the line are very well maintained for the differential visibility and somehow maintained for the differential phase. The global curvature of the differential phase is strongly affected, which looks similar to what we observe on the differential phase of 3C273. However, even when we simulate with exaggerated parameters, we do not seem able to « kill » a sharp differential phase feature of more than typically 3°. The simulation legitimates corrections of the « smooth » curvatures in the differential visibility and phase by a polynomial fit outside the emission line and outside the telluric lines.

The simulation also shows that when the source is « flat », i.e. when all chromatic OPD, window and detector biases are corrected from « models » in each frame before computing the 2DFTs, the tolerance to piston errors is much higher. This is currently being implemented in the data processing.

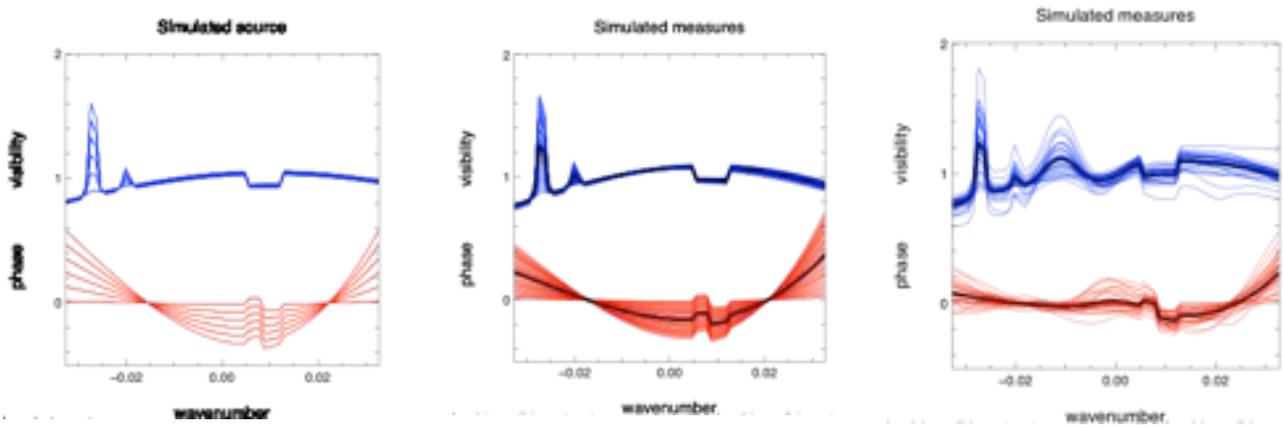

Figure 7: simulation of the « piston error bias » effect on the differential visibility and phase. Figure 7a (left): simulated source. In blue (top) the source differential visibility. In red (bottom) the source differential phase in radians. The chromatic OPD and the telluric line bias evolve like in the observations. Figure 7b (center): simulated measures with a piston rms=5 μm rms. The measures are almost unaffected. Figure 7c (right): simulated measures with a piston rms=50 μm rms. The measures are severely affected but sharp differential variations survive.

## 4. RESULTS AND DISCUSSION

Figure 8 shows the differential visibilities and phases obtained on 3C273 after calibration and our best current bias correction, based mainly on a polynomial fit of the measures outside the emission and absorption lines.

The differential visibility shows a drop in the emission line, which increases with the baseline. The various biases (and debiasing procedures) can create differential visibility artifacts and change the amplitude of the differential visibility drops, but cannot cancel them. From the simulation in the previous section, a very conservative error on our bias correction is of 0.02 per spectral channel. We finally have:

$$\frac{V_{line}(50m)}{V_{cont}(50m)} = 0.98 \pm 0.03, \quad \frac{V_{line}(80m)}{V_{cont}(80m)} = 0.94 \pm 0.04, \quad \frac{V_{line}(125m)}{V_{cont}(125m)} = 0.92 \pm 0.04$$

The errors per spectral bin *add linearly* the error estimated from statistical differences between the 47 exposures and the bias correction error. The spectral extension is of at least 2 spectral bins, i.e. 2500 km/s, which confirms that we are seeing BLR features.

About the differential phase, the most reasonable current statement is that there are no features larger than 3°, which corresponds to a photocenter displacement of 30 μas in the 125 m baseline direction.

This result is a surprise, which strongly stimulated our bias analysis. The BLR was supposed compact (<130 μas) and contained in the dust torus and thus the differential visibility was supposed to increase in the line (see figure 9). Let us give a very preliminary interpretation of our measures in simple geometric terms. All features observed in the K band continuum and in the emission line are small (less than 1 mas) with regard to the VLTI best resolution of 3.5 mas. Then each feature can be defined by its contribution to the total flux, its equivalent width and its photocenter shift, both in the direction of the projected observation baseline. The smaller scale details will not affect the measures at our baselines, beyond their impact on width, photocenter and flux. So we decide to represent the continuum image by a centered

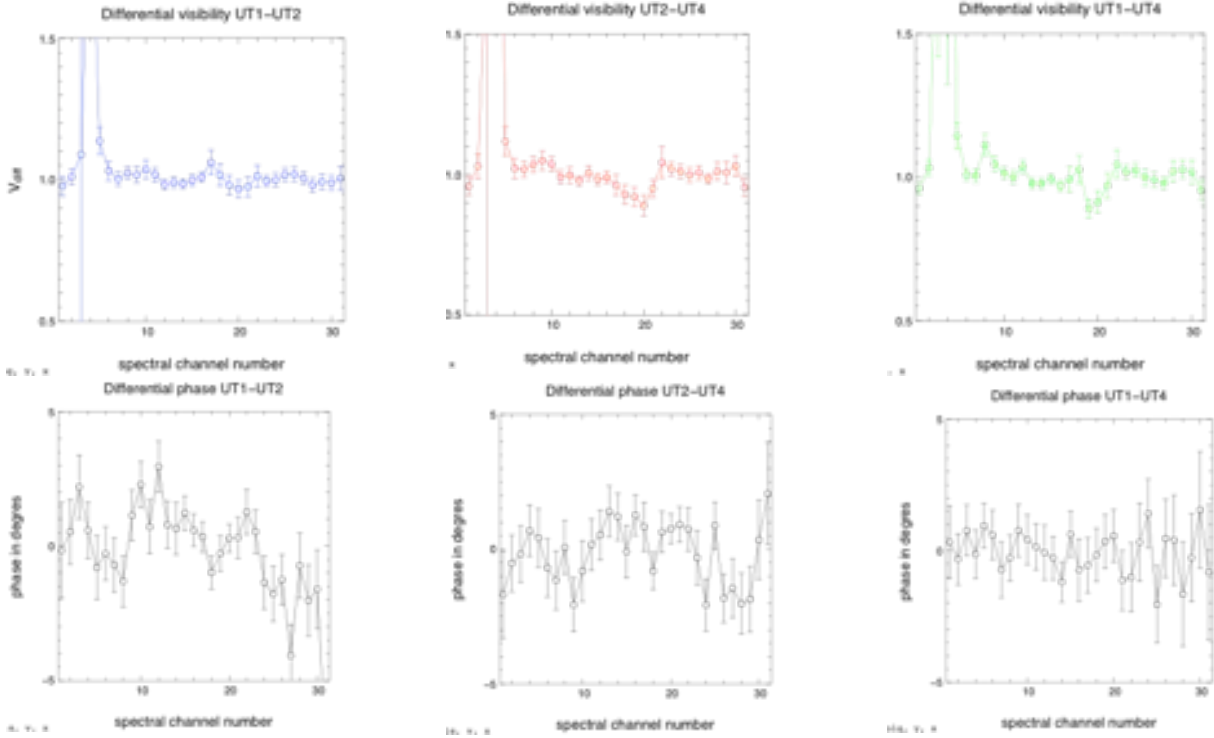

Figure 8: Provisional differential visibility (top) and phase (bottom) for a 50m (left), 80 m (center) and 125 m baselines. The differential visibility shows a drop in the emission line (see spectrum plot in figure 5a), which increases with the baseline. About the differential phase, the most reasonable statement is to say that there are no features larger than 2°. The error bars in the plots are estimated by the temporal dispersion over the 47 exposures.

Gaussian of FWHM adjusted to fit the absolute visibility measurements with the 84 m KI baseline[8]. This corresponds to a Gaussian with FWHM $= w_{cont} \approx 0.4\ mas$. The BLR is represented by a second Gaussian, with an intensity adjusted to fit the observed line profile over the continuum shown in figure 1. We change the width $w_{BLR}$ and the center $x_{BLR}$ of the BLR gaussian to try to fit our differential visibility measures. The results are displayed in figure 9. A centered compact BLR of angular size 0.13 mas, in agreement with the reverberation mapping + distance prediction, produces a differential visibility increase in the line of about 3%. To explain our visibility drops, we need either a very large centered BLR, with $w_{BLR} > 0.8\ mas$, or a compact (i.e. $w_{BLR} < 0.5\ mas$) BLR shifted by $x_{BLR} > 0.5\ mas$ with regard to the continuum photocenter. Qualitatively, an offset between the BLR and the continuum image makes sense. The continuum image contains contributions from various structures including the torus, with a possibly brighter face-on inner rim. There is actually no good reason for this image to be exactly centered on the accretion disk, except maybe for a face-on torus. Quantitatively, a shift $x_{BLR} > 0.5\ mas$ hardly fits in structure with an apparent radius $< 0.3\ mas$. But the critical difficulty of this interpretation is that the BLR shift would yield a photocenter displacement in the line of at least 250 µas in the direction of the baseline displaying the visibility drop, as shown in figure 9c. This corresponds to a differential phase of 25°, which does not appear in our differential phase measures with $|\phi(\lambda)| < 3°$ on all baselines. On all our baselines, the BLR FWHM is larger, or at least comparable to the diameter of the equivalent continuum ring, and the photocenter of the BLR coincides with this of the continuum image within 30 µas in the direction of the longest baseline (UT1-UT4) and 60 µas in the direction of the shortest one (UT1-UT2).

So, we must favor a centered and large BLR, of angular size larger than 0.8 mas, in the direction UT1-UT4 (60° $E \rightarrow N$), corresponding to a 0.4 mas radius of more than 1300 light days (if there is no major error on the distance of 3C273). This is in contradiction with the 300 ld RM radius observed in $H_\gamma$. Some investigation is needed to evaluate the possible ratio between the RM radii in $Pa_\alpha$ and $H_\gamma$. However, even if the RM radius can change with the line (it is $r = 514^{+65}_{-64}$ in $H_\alpha$), at least two types of arguments tend to discard a BLR much larger in $Pa_\alpha$ than in $H_\gamma$. First the $Pa_\alpha$ line profile is extremely similar to the $H_\beta$ in km/s. Second the higher energy transitions of $Pa_\alpha$ are usually more likely related to "inner" and hence "smaller" structures. Thus, we at least provisionally conclude that the 3C273 BLR has two different characteristic scales: the interferometric radius, larger than 1000 light days, and the RM radius, of the order of 300 light days. The angular visibility and the RM time delay correspond to integrals with different weightings of the different BLR parts. A detailed morphological modeling is needed to investigate their relative variation. It seems possible to imagine a BLR with two components: something compact around the accretion disk that dominates the RM, and is not resolved with our current differential phase accuracy, and a larger or shifted component, strong enough for a significant contribution to the angular intensity distribution, slow enough to explain that it has not been detected by spectroscopy and located in the right place, in the moment of our observations, to avoid introducing a photocenter shift in spite of its large size.

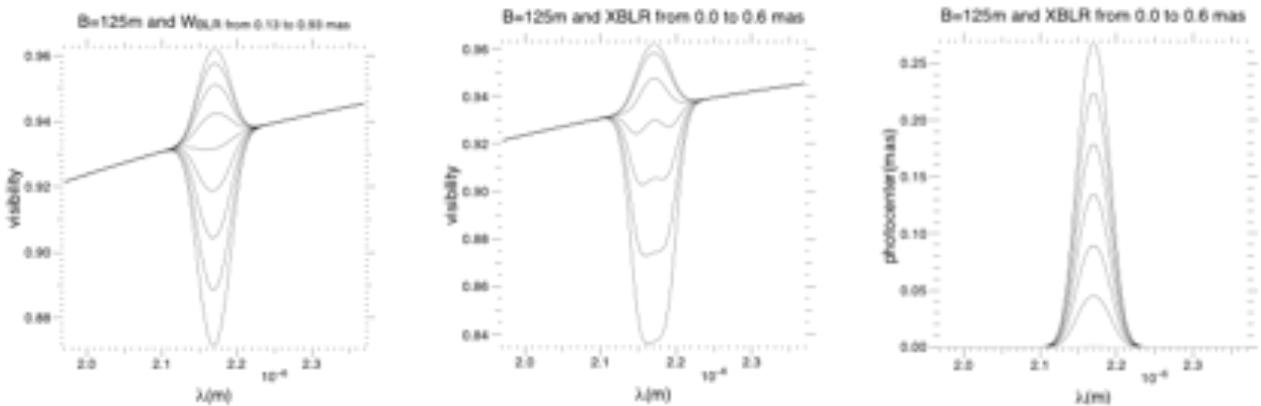

Figure 9: grid of « models » to interpret our differential visibility and phase measures. Left: visibility on a 125 m baseline for a centered BLR with widths ranging from 0.13 to 0.93 with 0.1 mas step (from top to bottom). Center, the same visibility for a compact 0.13 mas BLR shifted with regard to the continuum photocenter by a value ranging from 0 to 0.6 mas (top to bottom). Right: photocenter displacement in the baseline direction for the shifted BLR case (the photocenter displacement for the centered BLR is zero, and increases with the shift).

## 5. CONCLUSION

We have reported the first observations of the BLR of a quasar by optical interferometry. In medium spectral resolution, our so-called "blind mode" new observation technique allows a gain of at least two magnitudes with regard to the current fringe tracker. We believe that this must be used to revise the limiting magnitudes for spectrally resolved interferometric observations. Data processing based on the accumulation of 2D Fourier Transforms power spectra allowed an efficient fringe monitoring and slow coherencing on our K=10 target and we were probably at least one magnitude below the ultimate limit of the technique. The use of 2D FT cross spectra allowed to measure the differential visibility with an accuracy of 3% and a differential phase with an accuracy of about 3°. This allows measuring the source photocenter displacement with accuracy better than 30 micro arc seconds, and to constrain differential sizes smaller than 0.2 mas, to be compared with the standardly computed VLTI resolution of 3.5 mas.

We have analyzed the fundamental bias of this data processing technique, and taken it into account in the calibration of the data. Our current accuracy on the differential phase is limited by this bias correction. The data will be reprocessed with procedures less sensitive to the bias, and it is still possible to expect a final phase accuracy near 1°, in principle sufficient to resolve even compact 0.1 mas Keplerian structures.

The results show that the visibility of 3C273 is lower in the $Pa_\alpha$ emission line than in the K band continuum, and that the visibility drop increases with the baseline length, while the differential phase is smaller than 3° on all baselines. They imply that the BLR seen by interferometry is fairly large, with a radius of at least 1000 light days, comparable or slightly larger than the dust ring radius estimated from broadband low spectral resolution observations with the Keck interferometer. This result is quite surprising, mainly with regard to the 300 light days reverberation mapping radius. After a careful analysis of the data and of the possible biases, and even if this work is still in progress, we believe that a BLR with two different scales, and hence two different structures, the more than 1000 ld interferometric radius and the 300 ld reverberation mapping radius should be considered as a serious working hypothesis.

## ACKNOWLEDGEMENTS

The data has been obtained during the ESO run 087.B-0754 (A), an AMBER Consortium Guaranteed Time Observation. The authors thank Dr. Makoto Kishimoto, from the MPIfR, and Dr. Philippe Stee, from the Lagrange Laboratory, for the very useful discussions and comments.